\documentclass[
    ,final            
  ]
  {aipproc}

\layoutstyle{8x11single}

\begin{document}

\title{Excited-State Hadron Masses from Lattice QCD}


\author{C.~Morningstar}{
        address={Dept.~of Physics, Carnegie Mellon University, 
        Pittsburgh, PA 15213, USA}}

\author{J.~Bulava}{
        address={NIC, DESY, Platanenallee 6, D-15738, Zeuthen, Germany}}

\author{J.~Foley}{
        address={Physics Department, 
        University of Utah, 
        Salt Lake City, UT 84112, USA}}

\author{Y.C.~Jhang}{
        address={Dept.~of Physics, Carnegie Mellon University, 
        Pittsburgh, PA 15213, USA}}

\author{K.J.~Juge}{
        address={Dept.~of Physics, University of the Pacific, 
        Stockton, CA 95211, USA}}

\author{D.~Lenkner}{
        address={Dept.~of Physics, Carnegie Mellon University, 
        Pittsburgh, PA 15213, USA}}

\author{C.H.~Wong}{
        address={Dept.~of Physics, Carnegie Mellon University, 
        Pittsburgh, PA 15213, USA}}

\begin{abstract}
Progress in computing the spectrum of excited baryons and mesons 
in lattice QCD is described. Large sets of spatially-extended hadron 
operators are used.  The need for multi-hadron operators in addition
to single-hadron operators is emphasized,
necessitating the use of a new stochastic method of treating the
low-lying modes of quark propagation which exploits Laplacian Heaviside
quark-field smearing.  A new glueball operator is tested and 
computing the mixing of this glueball operator with a quark-antiquark
operator and multiple two-pion operators is shown to be feasible.
\end{abstract}

\maketitle


In a series of papers\cite{baryons2005A,baryons2005B,baryon2007,nucleon2009,
Bulava:2010yg,StochasicLaph}, we have been striving to 
compute the finite-volume stationary-state energies of QCD using Markov-chain
Monte Carlo integration of the QCD path integrals formulated on a
space-time lattice. Such calculations are very challenging.  
Computational limitations cause simulations to be done with quark masses 
that are unphysically large, leading to pion masses that are heavier 
than observed and introducing systematic errors in all other hadron energies.  
The use of carefully designed quantum field operators is crucial 
for accurate determinations of low-lying energies. To study a particular state 
of interest, the energies of all states lying below that state must first be 
extracted, and as the pion gets lighter in lattice QCD simulations, more and 
more multi-hadron states lie below the masses of the excited resonances.  The 
evaluation of correlations involving multi-hadron operators contains new 
challenges since not only must initial to final time quark propagation be 
included, but also final to final time quark propagation for a large number
of times must be incorporated.  The masses and widths of resonances must be 
deduced from the discrete spectrum of finite-volume stationary 
states for a range of box sizes.

To compute the QCD stationary-state energies, the matrices $C_{ij}(t)$ of 
temporal correlations of sets of single-hadron and multi-hadron operators are
estimated using the Monte Carlo method.  For an $N\times N$ matrix, the
$N$ eigenvalues of $C(t_0)^{-1/2}C(t)C(t_0)^{-1/2}$ tend to $\exp(-E_k (t-t_0))$
for large $t$ and fixed $t_0$, where the decay rates $E_k$ are the $N$ 
lowest-lying stationary-state energies that can be produced from the vacuum
by the operators used.  To compute the correlations involving isoscalar
mesons and good multi-hadron operators, quark propagation from all spatial
sites on one time slice to all spatial sites on another time slice 
are needed, and propagation from the sink time to the sink time are needed
for a large number of sink times.  A new method known as the stochastic
LapH method\cite{StochasicLaph} has been introduced to make such
computations accurate and practical in large volumes.

\begin{figure}[t]
  \includegraphics[width=6in,bb=0 15 567 299]{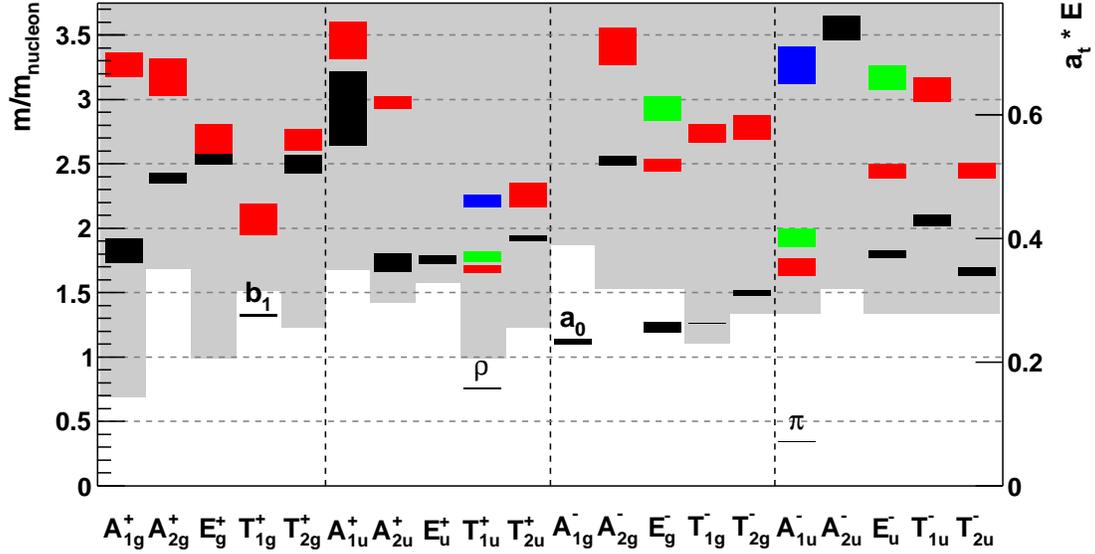}
  \caption{Masses of the isovector mesons in terms of the nucleon mass
using the stochastic LapH method with 170 gauge configurations on a $24^3\times 128$
anisotropic lattice.  The pion mass is about $m_\pi\sim 390$~MeV.
In the irrep labels, the letters with numerical subscripts refer
to the point group $O_h$ irreps, the subscripts  $g$ and $u$ refer to even and 
odd parity, respectively, and the superscripts
$\pm$ refer to $G$-parity.  Only single-hadron operators were used with dilution
scheme $(TF,SF,LI8)$.  The shaded region indicates the threshold locations for
multi-hadron energy levels.  We emphasize that extractions of energies in the
shaded regions could be complicated by ``false plateaux" unless multi-hadron
operators are used. \label{fig:isovectorspec}}
\end{figure}

Our first results for the isovector mass spectrum on a large $24^3\times 128$
anisotropic lattice are shown in Fig.~\ref{fig:isovectorspec}.  The pion mass
is about $m_\pi\sim 390$~MeV here.  These results are not finalized since
only single-hadron operators were used.  The shaded region indicates the threshold
locations for multi-hadron energy levels.  We emphasize that extractions of 
energies in the shaded regions could be complicated by ``false plateaux'' unless 
multi-hadron operators are used. The inclusion of the multi-hadron operators
is in progress.

\begin{figure}[b]
  \includegraphics[width=6in,bb=0 15 567 231]{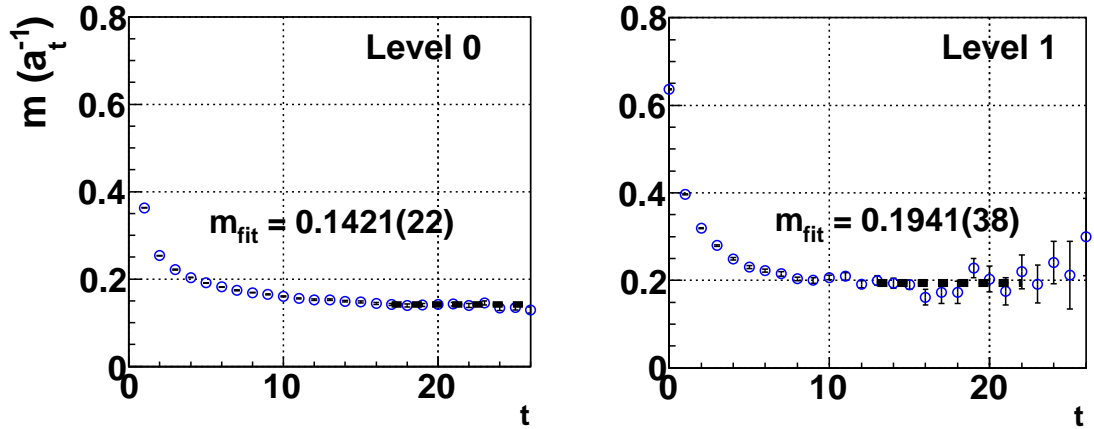}
  \caption{Effective masses corresponding to the diagonal elements of the rotated
$2\times 2$ correlator matrix involving a single-site $\rho$-meson operator and
an $I=1$ $\pi\pi$ operator in a $P$-wave with minimal relative momentum.  The $\rho$
operator dominates the lowest-lying level (left), while the $\pi\pi$ operator
dominates the first-excited state (right).  Although the mixing of the operators
is small, it is not negligible.  These results were obtained on 584 configurations
of the $24^3\times 128$ lattice with pion mass $m_\pi\sim 240$~MeV.
\label{fig:rhopipi}}
\end{figure}

Some initial results that incorporate multi-hadron operators are shown in
Fig.~\ref{fig:rhopipi}.  A $2\times 2$ correlation matrix was evaluated
involving a single-site $\rho$-meson operator and a total isospin $I=1$
$\pi\pi$ operator in a $P$-wave with minimal relative momentum. 
The stochastic LapH method enables very accurate
estimates of all elements (both diagonal and off-diagonal) of this
correlation matrix such that diagonalization can be done.  The effective
masses associated with the diagonalized correlator are shown in
Fig.~\ref{fig:rhopipi}. The $\rho$ operator dominates the lowest-lying level,
while the $\pi\pi$ operator dominates the first-excited state.  Although 
the mixing of the operators is small, it is not negligible.  This is 
certainly a warning about the dangers of extracting high-lying resonance
energies using only single-hadron operators.

\begin{figure}[t]
  \includegraphics[width=4in,bb=0 15 567 306]{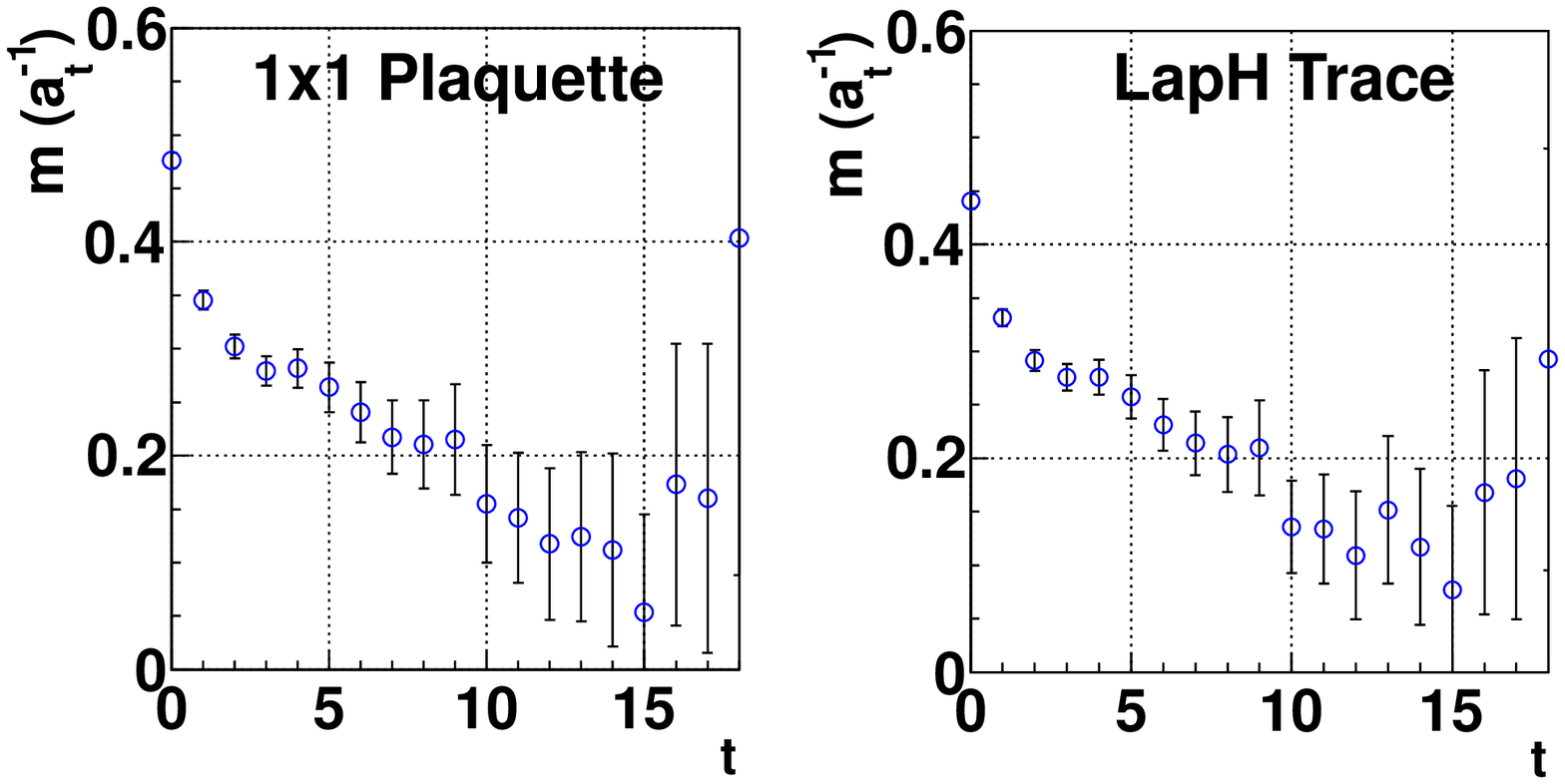}
  \caption{Comparison of the effective mass associated with the correlator
of the standard smeared plaquette glueball operator (left) with that of our
new glueball operator $G_\Delta$ defined in the text (right).  Similarity of the 
two results shows that the new glueball operator is just as useful as the
familiar smeared plaquette for studying the scalar glueball.  These effective
masses do not reach a plateau at the glueball mass since various $\pi\pi$
states and other multi-hadron states have smaller energies than the glueball mass.
These results were obtained using 584 configs of the $24^3$ ensemble for pion mass
$m_\pi\sim 240$~MeV. \label{fig:glueball}}
\end{figure}

Determining meson masses in the interesting scalar isoscalar sector
will ultimately involve including a scalar glueball operator, so we began
looking into the feasibility of such calculations.  LapH quark-field smearing
involves the covariant spatial Laplacian $\widetilde{\Delta}$.  The eigenvalues 
of the Laplacian
are invariant under rotations and gauge transformations so are appropriate
for a scalar glueball operator.  The lowest-lying eigenvalue was studied,
as well as other functions of the eigenvalues.  We found that any
combination of the low-lying eigenvalues worked equally well for
studying the scalar glueball.  In particular, the operator defined by
$G_\Delta(t) = 
-{\rm Tr}(\Theta(\sigma_s^2+\widetilde{\Delta})\widetilde{\Delta}(t))$
was used.  The effective mass associated with this operator is compared
to that of the smeared plaquette glueball operator in
Fig.~\ref{fig:glueball}.  Similarity of the results shows that $G_\Delta$
is just as useful as the familiar smeared plaquette for studying the scalar 
glueball. 

\begin{figure}[b]
  \includegraphics[width=6.3in,bb=0 15 567 155]{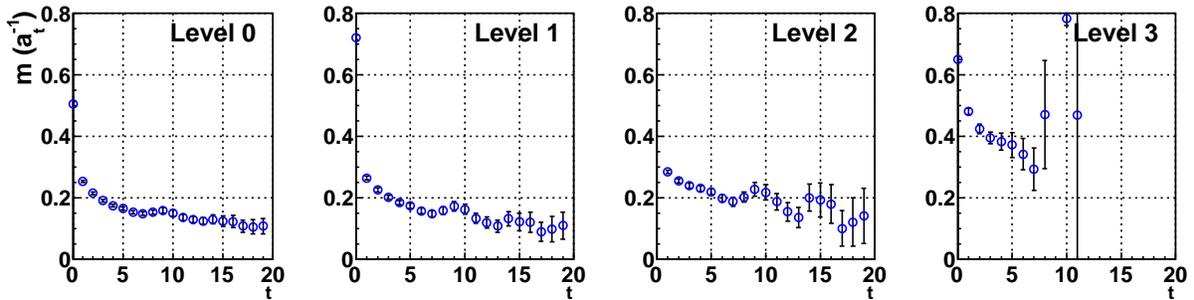}
  \caption{Effective masses corresponding to the diagonal elements of the rotated
$4\times 4$ correlator matrix involving a scalar isoscalar single-site 
quark-antiquark meson operator, the new glueball operator $G_\Delta$, and
two $I=0$ $\pi\pi$ operators in an $S$ wave (one with zero relative momentum and
the other with minimal nonzero relative momentum). Mixing of these operators is 
sizeable. More $\pi\pi$ operators must be included to reliably extract the glueball
mass.  These results were obtained on 100 configurations
of a $16^3\times 128$ lattice with pion mass $m_\pi\sim 390$~MeV.
\label{fig:isoscalar}}
\end{figure}

Effective masses corresponding to the diagonalized $4\times 4$ correlator 
matrix involving a scalar isoscalar single-site quark-antiquark meson operator, 
the new glueball operator $G_\Delta$, and two $I=0$ $\pi\pi$ operators in an 
$S$ wave (one with zero relative momentum and the other with minimal nonzero 
relative momentum) are shown in Fig.~\ref{fig:isoscalar}. Mixing of these 
operators is sizeable. More $\pi\pi$ operators must be included to reliably 
extract the glueball mass. 

These results demonstrate that the stochastic LapH method is useful for
accurately estimating all of the temporal correlations needed for a full
study of the QCD stationary-state energy spectrum, which we are currently
pursuing.  This work was supported by the U.S.~NSF
under awards PHY-0510020, PHY-0653315, PHY-0704171, PHY-0969863, and
PHY-0970137, and through TeraGrid/XSEDE resources provided by 
TACC and NICS under grant numbers TG-PHY100027 and TG-MCA075017.

\end{document}